\documentclass[10pt]{article}
\usepackage{amsmath, amssymb, latexsym, slashed, cancel, hyperref, accents}
\title{A realisation of Lorentz algebra in Lorentz violating theory}
\author{Oindrila Ganguly\footnote{oindrila@bose.res.in} \\
\textit{S. N. Bose National Centre for Basic Sciences, Kolkata 700098, India}}

\begin{document}
\maketitle

\begin{abstract}
A Lorentz non-invariant higher derivative effective action in flat spacetime, characterised by a constant vector, can be made invariant under infinitesimal Lorentz transformations by restricting the allowed field configurations. These restricted fields are defined as functions of the background vector in such a way that background dependance of the dynamics of the physical system is no longer manifest. We show here that they also provide a field basis for the realisation of Lorentz algebra and allow the construction of a Poincar\'e invariant symplectic two form on the covariant phase space of the theory.
\end{abstract}

\section{Introduction}

Any departure from exact Lorentz symmetry is expected to be a telltale footprint of quantum gravity, the theory of physics beyond the Planck scale (denoted by Planck mass $M_{Pl}$). This possibility is tremendously exciting owing to the fact that quantum gravity effects fall largely outside the domain of our current experiments and observations. Moreover, it is always challenging to find the limits of validity of any symmetry. These factors have stimulated a lot of work on the theoretical, phenomenological and experimental aspects of Lorentz violation since the last couple of decades \cite{carroll1989,latorre1995,colladay1996b,colladay1998,amelinocamelia1998gz,coleman1999ti,jacobson2000xp,colladay2001,sarkar2002mg,myers2003,jacobson2004qt,jacobson2005bg,mattingly2005re,maccione2009ju}. 

In the present report, we shall focus on a low energy effective field theory of scalar, vector and spinor fields developed by Myers and Pospelov \cite{myers2003} that incorporates deviation from Lorentz symmetry through the inclusion of an extra term. This new term contains a Planck suppressed dimension five operator with a constant background vector $\mathbf n$ that essentially assigns a preferred direction to spacetime. However, in \cite{og2011}, the authors demonstrate that there exist quite general field configurations which can restore Lorentz symmetry in the limit of infinitesimal transformations in modified action of \cite{myers2003}. These Lorentz preserving fields, as they have been named in \cite{og2011}, locally conserve the N\"other current for Lorentz transformations. Notice though that the Lorentz violation that had been woven into the Myers Pospelov theory is not lost. It manifests itself through the dependance of the Lorentz preserving fields on the directionality of the background i.e. on  $\mathbf n$. Given this scenario, a natural question to ask is whether Lorentz Lie algebra can be realised on a basis of such special field configurations.

We shall consider the case of the Lorentz preserving scalar fields \cite{og2011} for the modified action of  \cite{myers2003} as an illustrative example. The extended action of a complex scalar field $\phi$ proposed by Myers and Pospelov is,

\begin{align}
S_{MP_\phi} 
&= \int d^{4}x \mathcal{L}_{MP_\phi}~, \nonumber \\
&= \int d^{4}x \left[|\partial \phi|^2 - m^2|\phi|^2 \right] 
+ \int d^{4}x \frac{i\kappa}{M_{Pl}}\phi^*\partial_{n}^3\phi ~.
\label{eq:lmpscalar}
\end{align} 

The first integral contains the usual Lagrangian density of a complex scalar field with mass $m$ while the term under the second integral is the Lorentz violating contribution. $\kappa$ is a real, dimensionless parameter and $\mathbf{n}\cdot \mathbf{\partial} \equiv \partial_\mathbf{n}$, $\mathbf{n}$ being a constant four vector. According to \cite{og2011}, this action shall be symmetric under infinitesimal Lorentz boosts or rotations if the complex scalar field can be split as,

\begin{align}
\phi(\mathbf x) = \phi_{\parallel}(\mathbf{x}_{\parallel}) + \phi_{\perp}(\mathbf{x}_{\perp}) 
\label{eq:lpphi}
\end{align} 

where, $\phi_{\parallel}$ and $\phi_{\perp}$ are arbitrary functions of their
respective arguments $\mathbf{x}_{\parallel}$ and $\mathbf{x}_{\perp}$ defined by $\mathbf{n}$. $\mathbf{x}_{\parallel} \equiv \frac{\mathbf{x}\cdot \mathbf{n}}{n^2}\mathbf{n}$  and $\mathbf{n}\cdot \mathbf{x}_{\perp} = 0$, so that $\mathbf{x} = \mathbf{x}_{\parallel} + \mathbf{x}_{\perp}$. So, the derivative operator can be written as $\mathbf{\partial} = \mathbf{\partial}_{\mathbf x_\parallel} + \mathbf{\partial}_{\mathbf x_\perp} = \mathbf{\partial}_{\parallel} + \mathbf{\partial}_{\perp}$ where the notation $\mathbf{\partial}_{\parallel} \equiv \mathbf{\partial}_{\mathbf{x}_{\parallel}}$ and $\mathbf{\partial}_{\perp} \equiv \mathbf{\partial}_{\mathbf{x}_{\perp}}$.

A key aspect of the additional term in \eqref{eq:lmpscalar} is that it contains third order derivatives of the field unlike standard first order Lagrangians. But higher derivative Lagrangians are not new to physics \cite{podolsky1948}. Back in 1961, Ostrogradskii \cite{ostrogradskii1961} had developed a canonical formalism for dealing with them. Reviews and modifications of his technique  \cite{musicki1978qr,barcelosneto1989,simon1990,morozov2008} study mostly systems having finite number of degrees of freedom and higher time derivatives of the generalised coordinates. An extension to special relativistic continuous systems is presented in \cite{urries1998}, though it relies on the existence of Lorentz symmetry. In the next section, we shall briefly review Ostrogradskii's original construction and its generalisation to field systems, tailored to suit the Lagrangian \eqref{eq:lmpscalar} with fields that decouple as \eqref{eq:lpphi}

Usually, the canonical formalism is understood to be non-covariant because it involves the choice of a spacelike hypersurface and its orthogonal time direction. In fact, all the peculiarities of a higher derivative theory are due only to higher order time derivatives. Spatial derivatives are quite benign, staying within the scope of standard first order canonical approach. So, in section \ref{sec:mp}, the study of the modified scalar field theory will be split up into two cases distinguished by $\mathbf n^2$ being timelike and spacelike. We will not comment on $\mathbf n^2 = 0$ in this paper. We shall also take the liberty of working in certain Lorentz frames that simplify calculations. 

However, a covariant framework for the canonical formulation of a relativistic theory may also be developed  \cite{zuckerman1986,witten1986,300yofgr1987crnkovic,crnkovic1988,200yafterlag1991ashtekar} through the construction of a covariant phase space and a symplectic two-form on it. Reference \cite{aldaya1992} has presented a natural extension of this technique to higher derivative field theories. Section \ref{sec:covrev} will contain a summary of the main features of the covariant phase space for first order as well as higher order derivative field systems. We shall conclude by showing how the Lorentz preserving fields facilitate the construction of a covariant phase space structure for the Lorentz violating effective field theory under consideration.

\section{\label{sec:hd} Canonical formalism in the presence of higher derivatives} 

\subsection{\label{subsec:fdofhd} Non-relativistic systems with finite degrees of freedom} 

Let us consider a system described by the Lagrangian $L(q_a,d_t q_a,...,d_t^l q_a)$ which is a function not only of the generalised coordinate $q_a (t)$ and the velocity $d_t q_a(t)$ but of all derivatives $d_t q_a(t), d_t^2 q_a(t),...,d_t^l q_a(t)$ upto order $l$. Here $a$ labels the different degrees of fredom and we adopt the notation  $q_{a(j)} \equiv d_t^j q_a,\ j = 0,...,l$ so that $q_{a (j+1)} = \dot q_{a (j)}$. Each of these $q_{a(j)}$ upto $j = l-1$ will have a conjugate momentum $p^{a(j)}$. Remember though that the superscript $j$ of $p^{a(j)}$ only denotes that it is conjugate to $q_{a(j)}$. The system is now specified by a point in the phase space spanned by $q_{a(j)}, p^{a(j)}; j = 0,...,l-1$. We have here assumed that the highest derivative $q_{a(l)}$ can be written as a function of the other variables $q_{a(l)} (q_{a(0)}, p^{a(0)};...;q_{a(l-1)},p^{a(l-1)})$. The condition for extremisation of the action $S[q(t)] = \int dt\ L$ is,

\begin{align*}
\delta_q S = 0 
= \int dt \sum\limits_{j=0}^{l} (-d_t)^j \frac{\partial L}{\partial q_{a(j)}} \delta q_a 
+ \int dt d_t \sum\limits_{i=j+1}^{l} \sum\limits_{j=0}^{l-1}
(-d_t)^{i-(j+1)} \frac{\partial L}{\partial q_{a(i)}} \delta q_{a(j)}~.
\end{align*}

If the boundaries are such that the variations of $q_{a}$ and its derivatives upto order $l-1$ vanish, the second integral will go to zero and the equation of motion will be

\begin{align}
(-d_t)^j \frac{\partial L}{\partial q_{a(j)}} = 0~.
\label{eq:fdofeom}
\end{align}

On the other hand, if this system undergoes a symmetry transformation, $\delta S = 0$ and substitution of equation of motion \eqref{eq:fdofeom} yields,

\begin{align*}
d_t \left[\sum\limits_{i=j+1}^{l} \sum\limits_{j=0}^{l-1}
(-d_t)^{i-(j+1)} \frac{\partial L}{\partial q_{a(i)}} \delta q_{a(j)}\right] = 0~.
\end{align*}

The quantity within the square brackets is the conserved N\"other current $\mathcal J$. From its structure, we may read off the conjugate momenta

\begin{align}
p_{a}^{(j)} \equiv 
\sum\limits_{i=j+1}^{l}
(-d_t)^{i-(j+1)} \frac{\partial L}{\partial q_{a(i)}}~,
\ \ j=0,...,l-1~.
\end{align}

The N\"other current may then be cast into the standard form $\mathcal J = \sum\limits_{j=0}^{l-1} p_a^{(j)} \delta q_{a(j)}$.


\subsection{Relativistic continuous systems} \label{subsec:con}

Here, in place of the generalised coordinates, we shall be working with special relativistic fields having infinite number of degrees of freedom. Let us consider a system of scalar fields $\phi_a(\mathbf x)$. If $\mathcal{L} (\phi_a, \phi_{a, \rho_1}, \phi_{a, \rho_1 \rho_2},...,\phi_{a, \rho_1...\rho_l})$ be the Lagrangian density (where $\phi_{a, \rho_1...\rho_j} \equiv \partial_{\rho_1}...\partial_{\rho_j} \phi_a$), then by extremising the action $S[\phi(\mathbf x)] = \int d^4 x\ \mathcal{L}$ with respect to the fields $\phi_a(\mathbf x)$, we get the equation of motion

\begin{align}
\sum\limits_{j=0}^{l}
(-1)^j \partial_{\rho_1}...\partial_{\rho_j}
\frac{\partial \mathcal{L}}{\partial \phi_{a, \rho_1...\rho_j}} = 0 ~.
\label{eq:coneom}
\end{align}

The necessary boundary conditions are that the fields and their derivatives upto $\phi_{\rho_1...\rho_{l-1}}$ fall off at infinity. It would be convenient if we could write down a general form of the N\"other current for such higher derivative systems. This is achieved by applying a symmetry transformation to the fields and consequently employing the equation of motion \eqref{eq:coneom} to get the N\"other current,

\begin{align}
\mathcal{J}^{\rho_1}
= \sum\limits_{i=j+1}^{l} \sum\limits_{j=0}^{l-1} (-1)^{i-(j+1)}
\partial_{\rho_{j+2}}...\partial_{\rho_i}
\frac{\partial \mathcal{L}}{\partial \phi_{a, \rho_1...\rho_i}} \delta \phi_{\rho_2...\rho_{j+1}}~, 
\label{eq:ncgen} 
\end{align}

which is locally conserved i.e. $\partial_{\rho_1} \mathcal{J}^{\rho_1} = 0$. It may so happen that instead of $\delta \mathcal{L}$ being zero, the Lagrangian density varies by a total derivative. This would then contribute to the N\"other current.

However, as we have already mentioned, a canonical formalism of relativistic field theories requires us to foliate spacetime into spacelike hypersurfaces. This in turn involves the separation of temporal and spatial derivatives of the fields\footnote{At this stage, it is imperative that we sort the indices. Latin letters from the middle of the alphabet set viz. i,j,k,l are being used as summation indices while those from the end like r,s,...,z will denote spatial components. Different fields will be labelled by the alphabets a,b,c,d. Greek letters are reserved for spacetime indices.}. It is most often possible to arrange terms in the Lagrangian such that mixed derivatives of the fields as in $\partial_t^2 \partial_k \phi$ do not survive. This is true not only when the Lagrangian is Lorentz invariant \cite{urries1998} but also when the Lorentz violating action \eqref{eq:lmpscalar} is written in terms of the Lorentz preserving fields, as will be illustrated in the next section. In such situations, the canonical momenta will be given by

\begin{align}
\pi^{a(j)} \equiv 
\sum\limits_{i=j+1}^{l}
(-d_t)^{i-(j+1)} \frac{\partial \mathcal{L}}{\partial \phi_{a(i)}},
\ \ j=0,...,l-1~.
\label{eq:conmom}
\end{align}

The canonical variables will satisfy the Poisson bracket 

\begin{align}
\{\phi_{a (i)}(t, \vec x), \pi^{b (j)}(t, \vec{x}^{\prime})\} = \delta_a^b\ \delta_i^j\ \delta^{(3)}(\vec{x} - \vec{x}^{\prime})~.
\label{eq:conpb}
\end{align}


\section{Myers Pospelov theory} \label{sec:mp}

\subsection{Constant timelike background vector}

Without loss of generality, we are going to work in a Lorentz frame defined by $\mathbf{n} = (1,\vec 0)$. Then the Lorentz preserving fields $\phi_\parallel (t) , \phi_\perp (\vec x)$ become spatially homogeneous and static, spatially inhomogeneous respectively. This greatly simplifies the Lagrangian density:

\begin{align}
\mathcal{L}_{MP_\phi} 
&= \dot{\phi}_\parallel^*\dot{\phi}_\parallel - \vec{\nabla}\phi_\perp^* \cdot \vec{\nabla}\phi_\perp
+ \frac{i\kappa}{M_{Pl}}(\phi_\parallel^* + \phi_\perp^*)\dddot{\phi_\parallel} ~,
\label{eq:lagtmlk1}\\
&= \mathcal{L}_{MP_\phi} 
(\phi_\parallel,  \dot{\phi}_\parallel,  \ddot{\phi}_\parallel, \dddot{\phi}_\parallel,
\phi_\parallel^*, \dot{\phi}_\parallel^*, \phi_\perp, \vec{\nabla}\phi_\perp, 
\phi_\perp^*, \vec{\nabla}\phi_\perp^*) ~.
\label{eq:lagtmlk2}
\end{align}

Here, we have neglected the masses of the fields as we are interested in behaviour of the system at energies much higher than the field masses. It is now evident why our chosen Lorentz frame is particularly useful. Eq.\eqref{eq:lagtmlk1} has only higher order time derivatives of $\phi_\parallel (t)$ while $\vec{\nabla}\phi_\parallel = 0 = \partial_t \phi_\perp$. Thus all mixed derivatives in the sense described above will vanish. This permits us to safely use eq.\eqref{eq:conmom} to determine the canonical momenta. We list them in the following table. 

\begin{table}[h]
	\begin{center}
		\begin{tabular}{c l}
\hline
Generalised & Generalised \\
coordinate & momentum 
\\
\hline 
\vspace{-1ex}\\
$\phi_{\parallel (0)} = \phi_{\parallel}$ & $\pi_{\parallel (0)} = \dot{\phi}_{\parallel}^* 
						+ \frac{i\kappa}{M_{Pl}}
						\ddot{\phi}_{\parallel}^* $
\vspace{0.5ex}\\ 
$\phi_{\parallel (1)} = \dot{\phi}_{\parallel}$ &  $\pi_{\parallel (1)} = - \frac{i\kappa}{M_{Pl}}
						\dot{\phi}_{\parallel}^* $
\vspace{0.5ex}\\ 
$\phi_{\parallel (2)} = \ddot{\phi}_{\parallel}$ & $\pi_{\parallel (2)} 
						= \frac{i\kappa}{M_{Pl}}
						(\phi_\parallel^* + \phi_\perp^*)$
\vspace{0.5ex}\\ 
$\phi_{\parallel (0)}^* = \phi_{\parallel}^*$ & $\pi_{\parallel^* (0)} = \dot{\phi}_{\parallel}$
\vspace{0.5ex}\\ 
$\phi_{\perp (0)} = \phi_{\perp}$ & $\pi_{\perp (0)} = 0$
\vspace{0.5ex}\\ 
$\phi_{\perp (0)}^* = \phi_{\perp}^*$ & $\pi_{\perp^* (0)} = 0$ 
\\
		\end{tabular}
	\end{center}
\caption{Canonically conjugate phase space variables}
\label{tb:ntqp}
\end{table}

Note that $\pi_{\perp (0)} = 0$ and $\pi_{\perp^* (0)} = 0$ are constraint relations that will be imposed weakly. The equal time Poisson Bracket, eq. \eqref{eq:conpb}, will hold for the canonical variables.

Under an infinitesimal Lorentz transformation, $\delta_{\alpha \beta} \phi = x_{[\alpha} \partial_{\beta]} \phi$ and the Lagrangian density being a scalar function, also changes $\delta_{\alpha \beta} \mathcal{L} = x_{[\alpha} \partial_{\beta]} \mathcal{L} = \partial_{\rho_1}(x_{[\alpha} \delta^{\rho_1}_{\beta]}\mathcal{L})$. The Lorentz preserving fields conserve the N\"other current $\mathcal{J}^\mu_{\alpha \beta}$ \cite{og2011}. The N\"other charge is given by

\begin{align*}
Q_{\alpha \beta}
= \int_\Sigma d^3 \sigma_{\rho_1}\mathcal{J}^{\rho_1}_{\alpha \beta}~.
\end{align*}

Here, $\Sigma$ is a three dimensional hypersurface. If we orient it orthogonal to the time axis then,

\begin{align}
Q_{\alpha \beta}
&= \int_\Sigma d^3 \vec{x} 
\Big(
\pi_{\parallel}^{(0)} \delta_{\alpha \beta} \phi_{\parallel (0)}
+ \pi_{\parallel}^{(1)} \delta_{\alpha \beta} \phi_{\parallel (1)} 
+ \pi_{\parallel}^{(2)} \delta_{\alpha \beta} \phi_{\parallel (2)}
+ \pi_{\parallel^*} \delta_{\alpha \beta} \phi_{\parallel}^*
- x_{[\alpha} \delta^{0}_{\beta]}\mathcal{L}
\Big) ~,
\nonumber \\
&= \int_\Sigma d^3 \vec{x} 
\left( 
\pi_{a}^{(j)} \delta_{\alpha \beta} \phi_{a (j)}
- x_{[\alpha} \delta^{0}_{\beta]}\mathcal{L}
\right)~.
\label{eq:qntmlk2}
\end{align}

Here $\phi_{a} = \phi_{\parallel}, \phi_{\parallel}^*$. The other variables do not contribute. From the structure of $Q_{\alpha \beta}$ we can deduce that

\begin{align}
\{Q_{\alpha \beta} (t), \phi_{b (k)}(t, \vec x)\}
&= - \delta_{\alpha \beta} \phi_{b (k)}(t, \vec x), 
\ for \ \phi_b =  \phi_{\parallel}, \phi_{\parallel}^*~; \nonumber \\
&= 0 ,
\ for \ \phi_b =  \phi_{\perp}, \phi_{\perp}^*~. \label{eq:genntmlk}
\end{align}


The final step is to evaluate the algebra of the charges. A simple calculation using eq.,\eqref{eq:qntmlk2}, the basic Poisson bracket \eqref{eq:conpb} and the relation $\delta_{\alpha \beta} \phi = x_{[\alpha} \partial_{\beta]} \phi$ gives,

\begin{align}
\{Q_{\alpha \beta} (t), Q_{\rho \sigma} (t)\}
= \eta_{\alpha \sigma} Q_{\beta \rho} (t)
- \eta_{\alpha \rho} Q_{\beta \sigma} (t) 
+ \eta_{\beta \rho} Q_{\alpha \sigma} (t)
- \eta_{\beta \sigma} Q_{\alpha \rho} (t)~.
\label{eq:alntmlk}
\end{align}

Eqs. \eqref{eq:genntmlk}, \eqref{eq:alntmlk} confirm that N\"other charges defined in terms of the Lorentz preserving fields are generators of Lorentz transformation and satisfy the standard Lorentz algebra. Thus, the special field configurations provide a valid basis for the realisation of Lorentz Lie algebra which determines the local structure of Lorentz group near the identity.


\subsection{\label{subsec:splk} Constant spacelike background vector} 

Next, we go over to the Lorentz frame where $\mathbf{n} = (0, \vec 1)$. Then the Lorentz preserving fields will be $\phi_\parallel (\vec x) , \phi_\perp (t)$ and the Lagrangian will take the form

\begin{align}
\mathcal{L}_{MP_\phi} 
= \dot{\phi}_\perp^*\dot{\phi}_\perp - \vec{\nabla}\phi_\parallel^* \cdot \vec{\nabla}\phi_\parallel
+ \frac{i\kappa}{M_{Pl}}
(\phi_\parallel^* + \phi_\perp^*)
|\vec{\nabla}|^3 {\phi_\parallel}~.  \label{eq:lnsplk}
\end{align}

This Lagrangian density contains only first order time derivatives of the fields. Thus, no extra phase space variables will be required. The table below lists the pairs of canonical coordinates and momenta.

\begin{table}[h]
	\begin{center}
		\begin{tabular}{c l}
\hline
Generalised & Generalised \\
coordinate & momentum \\ 
\hline
\vspace{-1ex}\\
$\phi_{\parallel}$ & $\pi_{\parallel} = 0$
\vspace{0.5ex}\\ 
$\phi_{\parallel}^*$ &  $\pi_{\parallel ^*} = 0$
\vspace{0.5ex}\\ 
$\phi_{\perp}$ & $\pi_{\perp} = \dot{\phi}_\perp^*$
\vspace{0.5ex}\\ 
$\phi_{\perp}^*$ & $\pi_{\perp ^*} = \dot{\phi}_\perp$ \\
		\end{tabular}
	\end{center}
\caption{Canonically conjugate phase space variables for $\mathbf{n} = (0, \vec 1)$}
\label{tb:nsqp}
\end{table}

The fundamental Poisson Bracket is,

\begin{align}
\{\phi_{a}(t, \vec x), \pi_{b}(t, \vec{x}^{\prime})\} 
= \delta_{ab}\ \delta^{(3)}(\vec{x} - \vec{x}^{\prime})~.
\label{eq:pbns}
\end{align}

Integration of the zeroth component of the N\"other current over an appropriately chosen three dimensional spatial slice normal to the time axis gives the N\"other charge for Lorentz transformation,

\begin{align}
Q_{\alpha \beta}
= \int d^3 \vec{x} 
\left(
\pi_{\perp} \delta_{\alpha \beta} \phi_{\perp}
+ \pi_{\perp^*} \delta_{\alpha \beta} \phi_{\perp}^*
- x_{[\alpha} \delta^{0}_{\beta]}\mathcal{L}
\right) ~.
\label{eq:qnsplk} 
\end{align}

This is of the same form as the standard N\"other current obtained for a first order system of scalar fields. Hence, the N\"other charges would generate infinitesimal Lorentz transformation and satisfy the Lorentz algebra.


\section{Overview of covariant phase space formulation} \label{sec:covrev}

The covariant construction of phase space keeps intact Poincar\'e invariance of a physical system \cite{zuckerman1986,witten1986,300yofgr1987crnkovic,crnkovic1988,200yafterlag1991ashtekar}. As opposed to the standard decomposition of phase space in the $3+1$ framework, the classical covariant phase space $Z$ of a physical theory is defined as the space of classical solutions of the dynamical equations of the theory. Functions $\phi(\mathbf x)$, tangent vectors $\delta \phi$, one forms $\delta \phi (\mathbf x)$ and exterior derivatives $\delta$ can be defined on $Z$ following \cite{300yofgr1987crnkovic,crnkovic1988}.

The phase space $Z$ is naturally endowed with a closed, non-degenerate two-form $\tilde{\omega}$ (
$\tilde{\omega}$ is non-degenerate provided the one-form $\tilde{\omega} (\undertilde{V}) = 0$ if and only if $\undertilde{V} = 0$) called a symplectic structure. It can be written as an integral of some closed, conserved symplectic two-form current $\tilde{\omega}^\mu$ (note that $\tilde{\omega}^\mu$ is a two-form in phase space $Z$ and a vector current in spacetime; $\partial_\mu \tilde{\omega}^\mu = 0, \delta \tilde{\omega}^\mu = 0 $) over a hypersurface $\Sigma$,

\begin{align}
\tilde{\omega} = \int_{\Sigma} d\sigma_\mu \tilde{\omega}^\mu
\label{eq:symplcur}
\end{align} 

Hence, the task now is to find a suitable symplectic current, given any Lagrangian. Owing to our present interest in the covariant description of canonical formalism for higher derivative theories \cite{aldaya1992}, we shall straight away take the general example of a Lagrangian density $\mathcal{L} (\phi_a, \phi_{a, \rho_1}, \phi_{a, \rho_1 \rho_2},...,\phi_{a, \rho_1...\rho_l})$. For $\phi_{a}$ belonging to $Z$ and an arbitrary linear transformation $\delta \phi_a (\mathbf x)$ that takes $\phi_a (\mathbf x) \rightarrow \phi_a (\mathbf x) + \delta \phi_a (\mathbf x)$ on the phase space, we have 

\begin{align}
\delta \mathcal{L} 
&= \partial_\mu 
\sum\limits_{i=j+1}^{l} \sum\limits_{j=0}^{l-1} (-1)^{i-(j+1)}
\partial_{\rho_{j+2}}...\partial_{\rho_i}
\frac{\partial \mathcal{L}}{\partial \phi_{a, \mu\rho_2...\rho_i}} \delta \phi_{\rho_2...\rho_{j+1}} 
\nonumber \\
&= \partial_\mu j^\mu ,
\label{eq:presym}
\end{align}

after substituting the Euler Lagrange equation of motion \eqref{eq:coneom}. $j^\mu$ is interpreted as a pre-symplectic current because it is used to define the symplectic current $\tilde{\omega}^\mu = \delta j^\mu$. From eq. \eqref{eq:presym} one can see that it is obviously closed in phase space and conserved through its dependance on spacetime. Hence, the symplectic structure

\begin{align}
\tilde{\omega} = \int_{\Sigma} d\sigma_\mu \tilde{\omega}^\mu
= \delta \int_{\Sigma} d\sigma_\mu j^\mu
\label{eq:covsym}
\end{align}

is not only closed but also exact. Moreover, the local conservation of the symplectic current in spacetime guarantees that $\tilde{\omega}$ will not change with the choice of the surface of integration $\Sigma$ in \eqref{eq:covsym} and in particular, will be Poincar\'e invariant \cite{crnkovic1988}.

\section{Symplectic structure with Lorentz preserving fields} \label{sec:lpcov}

The covariant version of the canonical formalism is meaningful only when the physical theory has Poincar\'e invariance. The original Myers Pospelov Lorentz violating model doesn't meet this requirement. But when the allowed field configurations are restricted to the Lorentz preserving fields, the dynamics of the theory becomes invariant under infinitesimal Lorentz transformations enabling us to construct its covariant phase space. Only in this context, eq. \eqref{eq:presym} holds for $\mathcal{L}_{MP_\phi}$. 

One must have observed that the presymplectic current \eqref{eq:presym} and the N\"other current \eqref{eq:ncgen} have identical forms. In fact, with the interpretation of $\delta \phi_{a} (\mathbf x)$ as a one-form on phase space, the N\"other current becomes the pre-symplectic current one form. It must also be stressed that the entire construction of the symplectic structure \eqref{eq:covsym} follows from \eqref{eq:presym}, the validity of which is ensured by the Lorentz preserving fields. This is turn guarantees the existence of a Lorentz invariant (and Poincar\'e invariant) symplectic stucture on the covariant phase space of Lorentz preserving solutions of the equation of motion of Myers Pospelov action \cite{og2011}. 


\section*{Acknowledgements}
I am grateful to D. Gangopadhyay and P. Majumder for suggesting the problem and having numerous insightful discussions. I also thank A. V. Sleptsov and P. I. Dunin-Barkowski for pointing out reference \cite{duninbarkowski2008} to us.


\begin{thebibliography}{10}

\bibitem{carroll1989}
{\sc S.~M. Carroll}, {\sc G.~B. Field}, and {\sc R.~Jackiw},
\newblock {\em Phys.Rev.} {\bf D41}, 1231 (1990).

\bibitem{latorre1995}
{\sc J.~I. Latorre}, {\sc P.~Pascual}, and {\sc R.~Tarrach},
\newblock {\em Nuclear Physics B} {\bf 437}, 60  (1995).

\bibitem{colladay1996b}
{\sc D.~Colladay} and {\sc V.~A. Kostelecky},
\newblock {\em Phys.Rev.} {\bf D55}, 6760 (1997).

\bibitem{colladay1998}
{\sc D.~Colladay} and {\sc V.~A. Kostelecky},
\newblock {\em Phys.Rev.} {\bf D58}, 116002 (1998).

\bibitem{amelinocamelia1998gz}
{\sc G.~Amelino-Camelia}, {\sc J.~R. Ellis}, {\sc N.~Mavromatos}, {\sc D.~V.
  Nanopoulos}, and {\sc S.~Sarkar},
\newblock {\em Nature} {\bf 393}, 763 (1998).

\bibitem{coleman1999ti}
{\sc S.~R. Coleman} and {\sc S.~L. Glashow},
\newblock {\em Phys.Rev.} {\bf D59}, 116008 (1999).

\bibitem{jacobson2000xp}
{\sc T.~Jacobson} and {\sc D.~Mattingly},
\newblock {\em Phys.Rev.} {\bf D64}, 024028 (2001).

\bibitem{colladay2001}
{\sc D.~Colladay} and {\sc V.~A. Kostelecky},
\newblock {\em Phys. Lett.} {\bf B511}, 209 (2001).

\bibitem{sarkar2002mg}
{\sc S.~Sarkar},
\newblock {\em Mod.Phys.Lett.} {\bf A17}, 1025 (2002).

\bibitem{myers2003}
{\sc R.~C. Myers} and {\sc M.~Pospelov},
\newblock {\em Phys.Rev.Lett.} {\bf 90}, 211601 (2003).

\bibitem{jacobson2004qt}
{\sc T.~Jacobson}, {\sc S.~Liberati}, and {\sc D.~Mattingly},
\newblock {\em Springer Proc.Phys.} {\bf 98}, 83 (2005).

\bibitem{jacobson2005bg}
{\sc T.~Jacobson}, {\sc S.~Liberati}, and {\sc D.~Mattingly},
\newblock {\em Annals Phys.} {\bf 321}, 150 (2006).

\bibitem{mattingly2005re}
{\sc D.~Mattingly},
\newblock {\em Living Rev.Rel.} {\bf 8}, 5 (2005).

\bibitem{maccione2009ju}
{\sc L.~Maccione}, {\sc A.~M. Taylor}, {\sc D.~M. Mattingly}, and {\sc
  S.~Liberati},
\newblock {\em JCAP} {\bf 0904}, 022 (2009).

\bibitem{og2011}
{\sc O.~Ganguly}, {\sc D.~Gangopadhyay}, and {\sc P.~Majumdar},
\newblock {\em Europhys.Lett.} {\bf 96}, 61001 (2011).

\bibitem{podolsky1948}
{\sc B.~Podolsky} and {\sc P.~Schwed},
\newblock {\em Rev. Mod. Phys.} {\bf 20}, 40 (1948).

\bibitem{ostrogradskii1961}
{\sc M.~V. Ostrogradskii},
\newblock {\em Complete Collected Works}, volume~2,
\newblock Akad. Nauk Ukrain, SSR, Kiev, 1961,
\newblock in Russian.

\bibitem{musicki1978qr}
{\sc D.~Musicki},
\newblock {\em J.Phys.A} {\bf A11}, 39 (1978).

\bibitem{barcelosneto1989}
{\sc J.~Barcelos-Neto} and {\sc N.~R. Braga},
\newblock {\em Acta Phys.Polon.} {\bf B20}, 205 (1989).

\bibitem{simon1990}
{\sc J.~Z. Simon},
\newblock {\em Phys.Rev.} {\bf D41}, 3720 (1990).

\bibitem{morozov2008}
{\sc A.~Morozov},
\newblock {\em Theor.Math.Phys.} {\bf 157}, 1542 (2008).

\bibitem{urries1998}
{\sc F.~de~Urries} and {\sc J.~Julve},
\newblock {\em J.Phys.A} {\bf A31}, 6949 (1998).

\bibitem{zuckerman1986}
{\sc G.~J. Zuckerman},
\newblock {\em Conf. Proc.} {\bf C8607214}, 259 (1986).

\bibitem{witten1986}
{\sc E.~Witten},
\newblock {\em Nucl.Phys.} {\bf B276}, 291 (1986).

\bibitem{300yofgr1987crnkovic}
{\sc C.~Crnkovic} and {\sc E.~Witten},
\newblock {\em {Covariant description of canonical formalism in geometrical
  theories}}, chapter~16, pp. 676--684,
\newblock Cambridge University Press, Cambridge, 1987.

\bibitem{crnkovic1988}
{\sc C.~Crnkovic},
\newblock {\em Class.Quant.Grav.} {\bf 5}, 1557 (1988).

\bibitem{200yafterlag1991ashtekar}
{\sc A.~Ashtekar}, {\sc L.~Bombelli}, and {\sc O.~Reula},
\newblock {\em {The covariant phase space of asymptotically flat gravitational
  fields}}, pp. 417--450,
\newblock Elsevier Science Publishers B. V., North-Holland, Amsterdam
  (Netherlands), 1991.

\bibitem{aldaya1992}
{\sc V.~Aldaya}, {\sc J.~Navarro-Salas}, and {\sc M.~Navarro},
\newblock {\em Phys.Lett.} {\bf B287}, 109 (1992).

\bibitem{duninbarkowski2008}
{\sc P.~Dunin-Barkowski} and {\sc A.~Sleptsov},
\newblock {\em Theor.Math.Phys.} {\bf 158}, 61 (2009).


\end{thebibliography}
\end{document}